\documentclass[epsfig,psfig,amsmath]{article}
\usepackage{epsfig}
\def\theequation{\arabic{section}.\arabic{equation}}
\def\r{r^2_0}
\begin{document}
\def\bbox{\vrule height2mm depth0mm width5pt}
\def\theequation{\arabic{equation}}
\font\sqi=cmssq8
\def\DR{\rm I\kern-1.45pt\rm R}
\def\DH{\rm I\kern-1.45pt\rm H}
\def\DC{\kern2pt {\hbox{\sqi I}}\kern-4.2pt\rm C}

 \pagenumbering{arabic}
\setcounter{page}{1}
\newcommand{\beq}{\begin{equation}}
\newcommand{\eeq}{\end{equation}}
\newcommand{\bea}{\begin{eqnarray}}
\newcommand{\eea}{\end{eqnarray}}
\newcommand{\bc}{\begin{center}}
\newcommand{\ec}{\end{center}}
\newcommand{\mb}{\mbox{\ }}
\newcommand{\bs}{\mbox{\boldmath $\sigma$}}
\newcommand{\bp}{\mbox{\boldmath $\pi$}}
\def\r{r^2_0}
\newcommand{\ch}{{\tt h}}
\newcommand{\ra}{\rightarrow}
\newcommand{\la}{\leftarrow}
\newcommand{\IR}{\mbox{I \hspace{-0.2cm}R}}
\newcommand{\ol}{\overline}
\newcommand{\ul}{\underline}
\begin{center}
{\large \bf Quantum Oscillator on $\DC P^n$ in a constant magnetic field}\\
\vspace{0.5 cm}
{ \large Stefano Bellucci$^1$,    Armen Nersessian$^{2,3}$ and
Armen Yeranyan$^2$ }
\end{center}
{\it $^1$ INFN-Laboratori Nazionali di Frascati,
 P.O. Box 13, I-00044, Frascati, Italy\\
$^2$ Yerevan State University, Alex  Manoogian St., 1, Yerevan,
375025, Armenia\\
$^3$ Yerevan Physics Institute, Alikhanian Brothers St., 2, Yerevan, 375036,
 Armenia}
\begin{abstract}
\noindent
We construct  the quantum oscillator interacting with a constant magnetic field
 on complex projective spaces $\DC P^N$,
as well as on their non-compact counterparts, i. e.
the $N-$dimensional Lobachewski spaces ${\cal L}_N$.
We find the spectrum of this system and the complete basis of wavefunctions.
Surprisingly,
 the inclusion of a magnetic field does not yield any qualitative change
 in the energy spectrum.
For $N>1$ the magnetic field does not break the superintegrability of the system, whereas
for $N=1$ it preserves the exact solvability of the system.
 We  extend this results to
the cones constructed over $\DC P^N$ and ${\cal L}_N$,
 and
perform the  (Kustaanheimo-Stiefel) transformation of these systems
to the three-dimensional  Coulomb-like systems.

\end{abstract}
\setcounter{equation}0
\noindent
\subsubsection*{Introduction}
The harmonic oscillator plays a fundamental role in quantum mechanics.
 On the other hand, there are few articles related with
the oscillator on curved spaces.
The most known generalization of the Euclidian oscillator
is the oscillator on curved spaces with constant curvature
(sphere and hyperboloid) \cite{higgs} given by the potential
\beq
V_{Higgs}=\frac{\omega^2\r}{2}\frac{{\bf x}^2}{x^2_0},
\qquad \epsilon{\bf x}^2+ x^2_0=\r ,\quad \epsilon=\pm 1.
\eeq
This system received much attention since
its introduction (see for a review \cite{pogos} and refs. therein) and
 is presently known under the name of ``Higgs oscillator''.

Recently the generalization of the oscillator to  K\"ahler spaces has also
been suggested, in terms of the potential
\cite{cpn}
 \beq
V_{osc}=
\omega^2 g^{\bar a b}\partial_{\bar a}K \partial_b K . \label{op} \eeq
Various properties of the systems with this
 potential were studied in Refs. \cite{cpn,ny,bny,sqs}.
It was shown that on  the complex projective spaces
$\DC P^N$ such a system inherits
the whole set of rotational symmetries and a part of the hidden symmetries of
the $2N-$dimensional flat oscillator \cite{cpn}.
In Ref. \cite{ny}, the classical solutions of the system
on $\DC P^2$, ${\cal L}_2$ (the noncompact counterpart of $\DC P^2$)
and the related cones were presented, and the reduction to
three dimensions was studied.
Particulary, it was found that the oscillator
on some  cone related with $\DC P^2$ (${\cal L}_2$) results,
after Hamiltonian reduction, in the Higgs oscillator on the three-dimensional
sphere (two-sheet hyperboloid) in the presence of a Dirac monopole field.
In Ref. \cite{bny} we presented the exact quantum mechanical solutions for the
oscillator on $\DC P^2$, ${\cal L}_2$ and related cones.
We also reduced  these quantum systems to three dimensions and
performed their (Kustaanheimo-Stieffel) transformation to the three-dimensional
Coulomb-like systems. The ``K\"ahler oscillator'' is a distinguished
system with respect to supersymmetrisation as well.
Its preliminary studies were presented in \cite{sqs,cpn}.

In this paper we  present
 the exact solution of the quantum oscillator  on arbitrary-dimensional
 $\DC P^N$, ${\cal L}_N$ and related cones
   {\it in the presence of a constant magnetic field}.
The study of such systems is not merely of academic interest.
It is also relevant to  the  higher-dimensional quantum
 Hall effect. This theory has been formulated
 initially on the four-dimensional sphere \cite{zhang} and further
 included,
as a particular case, in the theory of the
quantum Hall effect on complex projective spaces
\cite{kn} (see, also \cite{qheother}).
 The latter theory is based on the quantum mechanics on $\DC P^N$
in a constant magnetic field.
Our  basic observation  is that the
inclusion of the constant magnetic field does not break
any existing hidden symmetries of the  $\DC P^N$-oscillator
and, consequently, its  superintegrability and/or exact solvability are preserved.

To be more concrete, let us consider
first the (classical) oscillator on $\DR^{2N}=\DC^N$.
It is described by the symplectic structure
\beq
\Omega_0=
d\pi_a\wedge dz^a+ d{\bar\pi}_a\wedge d{\bar z}^a \label{2}
\eeq
and the Hamiltonian
\beq
{\cal H}=\pi {\bar \pi}+\omega^2 z\bar z .\label{3}
\eeq
It has a symmetry group $U(2n)$ given by the generators of $SO(2n)$ rotations
\beq
J^+_{ab}={\bar z}^a\pi_b,\quad J^-_{\bar a\bar b}={ z}^a\bar\pi_b,
\quad J_{a\bar b}=iz^b\pi_a
-i\bar\pi_b\bar z^a
\eeq
and the hidden symmetries
\bea
&&I^+_{ab}=\pi_a\pi_b+\omega^2\bar z^a \bar z^b,
\quad I^-=
\bar\pi_a
\bar\pi_b+\omega^2     z^a      z^b
          ,\label{4}\\
&&I_{a\bar b}=\pi_a\bar\pi_b+\omega^2\bar z^a  z^b.\label{5}
\eea
 The oscillator on the complex
projective space $\DC P^N$ (for $N>1$) \cite{cpn}  and  the one on Lobacewski space ${\cal L}_N$
are defined by the same symplectic structure as above, see eq. (\ref{2}), with the Hamiltonian
\beq
 {\cal H}=g^{\bar a b}\bar\pi_a\pi_b+\frac{\omega^2{\r}}{2} z\bar z,
 \qquad g^{\bar a b}=\frac{2}{\r}(1+\epsilon z\bar
 z)(\delta^{ab}+\epsilon z^a{\bar z}^b),\qquad \epsilon=\pm 1.
\label{31}\eeq
The choice $\epsilon =1$ corresponds to $\DC P^N$, and $\epsilon=-1$ is associated to ${\cal
 L}_N$.
This  system inherits only part of the rotational and hidden symmetries
 of the  $\DC^N$-oscillator
given, respectively, by the following constants of motion:
\beq
J_{a\bar b}={i}(z^b\pi_a-\bar\pi_b\bar z^a), \quad
I_{a\bar b}=
2\frac{J^+_a J^-_b}{\r} +\frac{\omega^2\r}{2} {\bar z}^a z^b\; ,
\label{symc}\eeq
where $J^+_a={i}\pi_a+{i}\epsilon (\bar z\bar\pi)\bar z^a$,
 $J_a^-=\bar J^+_a$ are the translation generators.
It is clear that $J_{a\bar b}$  defines the $U(N)$ rotations,
 while  $I_{a\bar b}$ is just a $\DC P^N$ counterpart of
(\ref{5}).\\
In order to include a constant magnetic
 field, we have to leave
 the initial Hamiltonian unchanged,
  and replace  the initial symplectic structure
 (\ref{2}) by the following one:
\beq
\Omega_B=\Omega+iBg_{a\bar b}dz^a\wedge d\bar z^b,
\label{ss}\eeq
where $g_{a\bar b}$ is a
K\"ahler
metric of the  configuration space.
It is easy to  observe, that
 the inclusion of a constant magnetic field
preserves only the  symmetries  of the  $\DC^N$-oscillator
 generated
by $J_{a\bar b}$ and $I_{a\bar b}$. On the other hand,
the inclusion of the magnetic field
preserves all the symmetries for the oscillator on $\DC P^N$.
Hence, in the presence of a magnetic field the oscillators
on $\DC^N$ and $\DC P^N$ look much more similar,
than in its absence
\footnote{Notice, that in  \cite{cpn}, because of  arithmetic  mistake
in calculations, it was wrongly stated
that the inclusion of a constant magnetic field breaks the hidden symmetries
of the oscillator on $\DC P^N$.}.
Hence we could be sure
that the  $\DC P^N$- oscillator
preserves its classical and quantum exact solvability
in the presence of a constant magnetic field.

Below,
 we formulate  the Hamiltonian  and quantum-mechanical systems,
describing the $\DC P^N$-  and ${\cal L}_N$- oscillators
in the constant magnetic field and present their wavefunctions and spectra.
We also extend these results to the cones and discuss some related topics.

\subsubsection*{Oscillator in
a constant magnetic field: $\DC P^N$ and ${\cal L}_N$}
The classical K\"ahler oscillator in a constant magnetic field
is defined by the Hamiltonian
\beq
{{\cal H}}=g^{a\bar b}{\pi}_{a}{\bar\pi}_b +
\omega^2 g^{\bar a b}K_{\bar a}  K_b \quad ,
\label{qhc}\eeq
and the Poisson brackets corresponding to the symplectic structure
(\ref{ss})
\beq
\{\pi_a,z^b\}=\delta^b_a,\quad
\{\bar \pi_a,\bar z^b\}=\delta^b_a, \quad\{\pi_a,\bar\pi_b\}=iBg_{a\bar b}.
\label{hn}\eeq
Its
canonical quantization assumes the following choice of momenta operators:
\beq
\widehat{\pi}_a=-i\left(\hbar\partial_a+ B K_a/2 \right),\quad
\widehat{\bar\pi}_a=-i\left(\hbar\partial_{\bar a}-BK_{\bar a}/2 \right),
\eeq
where $\partial_a=\partial/\partial z^a$, $\partial_{\bar a}=
\partial/\partial {\bar z}^a$, $K_a=\partial K/\partial {z}^a$,
$K_{\bar a}=\partial K/\partial {\bar z}^a$.
\footnote{It is easy to see, that these operators are not
Hermitian with respect to the scalar product
$$
\left(\Psi,\Phi \right)=\int \bar\Psi(z, \bar z),\Phi(z,\bar z)  g [dz d\bar z],\quad g\equiv \det g_{a\bar b}$$.
In order to make the momenta operators Hermitian, we have to replace them as follows:
$$
\widehat{\pi}_a\to \widehat{\pi}_a- \frac{i\hbar} {2} \frac{\partial g }{\partial z^a},
\widehat{\bar\pi}_a\to \widehat{\bar\pi}_a-\frac{i\hbar} {2} \frac{\partial g }{\partial\bar z^a}.
$$ }
The quantum Hamiltonian looks  similar to the
classical one
\beq
\widehat{{\cal H}}=\frac{1}{2}g^{a\bar b}(\widehat{\pi}_{a}\widehat{\bar\pi}_b+\widehat{\bar\pi}_b\widehat{\pi}_{a}) +
\omega^2 g^{\bar a b}K_{\bar a}  K_b .\quad
\label{qh0}\eeq
In the specific case
of the complex projective space
$\DC P^N$ and its noncompact version, i.e. the Lobacewski space  ${\cal L}_N$,
we have to choose
  \begin{equation}
K=\frac{r^2_0}{2\epsilon}\log (1+ \epsilon z\bar z), \qquad
K_a=\frac{\r}{2\epsilon}\frac{\bar z^a}{1+\epsilon z\bar z},\quad
K_{\bar a}=\frac{\r}{2\epsilon}\frac{z^a}{1+\epsilon z\bar z}.
\eeq
The scalar curvature $R$  is related with the
parameter $\r$ as follows: $R=2\epsilon N(N+1)/\r$.
These systems  possess the $u(N)$ rotational symmetry  generators
\beq
\widehat{J}_{a\bar b}={i}\epsilon(z^b{\widehat \pi}_a-
\bar z^a{\widehat{\bar\pi}}_b )+
{i}\epsilon (z^c\widehat\pi_c-
\bar z^c \widehat{\bar\pi}_c
)\delta^{\bar a b} -
 \frac{B\r }{2}\frac{\epsilon{\bar z}^a z^b-
\delta^{\bar a b}}{1+\epsilon z\bar z},
\eeq
and the hidden symmetry defined by the generators
 \beq
{\widehat I}_{a\bar b}=
\frac{\widehat{{J}}^+_a\widehat{{J}}^-_b
+\widehat{{J}}^-_b\widehat{{J}}^+_a}{\r} +
\frac{\omega^2\r}{2} {\bar z}^a z^b\; ,
\label{sym}\eeq
where
$\widehat{{J}}^\pm_{a}$ are the translations generators
\beq
\widehat{{J}}^{+}_a=i\widehat\pi_a+
i\epsilon\bar z^a(\bar z\widehat{\bar\pi})-
\frac{B\r}{2}\frac{{\bar z}^a}{1+\epsilon z\bar z},
\quad
\widehat{{J}}^{-}_a=
-i\widehat{\bar\pi}_a-
i\epsilon z^a( z\widehat{\pi})-
\frac{B\r}{2}\frac{{ z}^a}{1+\epsilon z\bar z}.
\eeq
The Hamiltonian
 (\ref{qh0}) could be rewritten as follows:
\bea
{\cal H}=\frac{\hbar^2}{2\r}\left[-\frac{(1+\epsilon
x^2)^{N+1}}{x^{2N-1}}\frac{\partial}{\partial
x}\left(\frac{x^{2N-1}}{(1+\epsilon
x^2)^{N-1}}\frac{\partial}{\partial x}\right)
+\frac{4(1+\epsilon x^2)}{\epsilon x^2}{\widehat{\bf J}}^2
+{\epsilon}(1+\epsilon x^2)({2\widehat{J}_0}+
\frac{\mu_B}{\epsilon})^2\right]+\nonumber\\
+\frac{\omega^2\r x^2}{2}-\epsilon\frac{\hbar^2\mu_B^2}{2\r},
\eea
where
\beq
x=|z|, \quad\mu_B=\frac{B\r}{2\hbar},\quad 2\widehat{J}_0=z^a\partial_a-\bar z^{\bar a} \partial_{\bar a},
  \eeq
$\widehat{\textbf{J}}^2$ is the quadratic Casimir of
 the $SU(N)$  momentum operator  for the  $N>1$,
 and $\widehat{\textbf{J}}=\widehat{J}_0$  for $N=1$.
In order to get  the energy spectrum of the system, let us
consider the spectral problem
\beq
\widehat{{\cal H}}\Psi=E\Psi,\qquad
\widehat{J}_0\Psi=s\Psi ,
\qquad\widehat{\bf J}^2\Psi =j(j+N-1)\Psi.
\eeq
It is convenient to pass to the $2N-$dimensional spherical coordinates
$(x,\phi_i)$,
 where $i=1,\ldots, 2N-1$,    $x$ is a dimensionless radial coordinate
taking values in the interval  $ [0,\infty)$  for $\epsilon=+1$,
and in $[0, 1]$ for $\epsilon=-1$, and $\phi_i$ are appropriate
angular coordinates. The convenient
algorithm for the expansion of ``Cartesian'' coordinates to the spherical
ones is described in Appendix.
In the new coordinates the above system could be solved by the following choice
of the wavefunction:
\beq
\Psi=\psi(x) D^j_{s}(\phi_i)\label{sep},\eeq
where
$ D^j_{s}(\phi_i)$  is the eigenfunction of the
operators  $\widehat{\bf J}^2$, $\widehat J_0$.
It could be explicitly expressed  via $2N-$dimensional  Wigner functions,
 $ D^j_{s}(\phi_i) =
\sum_{m_i} c_{m_i}D^j_{\{m-i\},s}(\phi_i)$,
where
$j$, $m_i$ denote total and azimuth angular momenta,
 respectively, while
$s$ is the eigenvalue of the operator $\widehat{J}_0$
\beq
 m_i ,s=-j,-j+1,\ldots , j-1, j\;\;  j=0,1/2,1,\ldots
\label{jsm}
\eeq
 Now, we make the  substitution
\beq
x=\frac{\tan[\sqrt{\epsilon}\theta]}{\sqrt{\epsilon}}, \quad
\psi=\frac{f[\theta]\epsilon^{N/2-1/4}}{(\sin[\sqrt{\epsilon}\theta])^{N-1/2}(\cos[\sqrt{\epsilon}\theta])^{1/2}}
\label{sub}
\eeq
which yields the  the following  equation:
 \beq
f''+\epsilon\left(\tilde{E}-\frac{j^2_1-1/4}{\sin^2[\sqrt{\epsilon}\theta]}-
\frac{\delta^2-1/4}{\cos^2[\sqrt{\epsilon}\theta]}\right)f=0, \label{pt}\eeq
where \beq
\tilde{E}=\frac{2E\r}{\epsilon\hbar^2}+\frac{\omega^2r_0^4}{\hbar^2}+
N^2+{\mu_B}^2,
\qquad \delta^2=\frac{\omega^2r_0^4}{\hbar^2}+
\left(2s+\frac{\mu_B}{\epsilon}\right)^2, \qquad
j_1=2j+N-1.
\label{te}\eeq
The   regular wavefunctions, which form a complete
orthonormal  of the above Schroedinger equation, are of the form
 \beq
\psi=\left\{
\begin{array}{c}
C\sin^{j_1-1}\theta\cos^{\delta}\theta\; _2F_1(-n,n+\delta+j_1+1;j_1+1;\sin^{2}\theta),\;{\rm for \;} \epsilon=1\;\;\\
C\sinh^{j_1-1}\theta\cosh^{-\delta+2n}\theta\; _2F_1(-n,-n+\delta,j_1+1,\tanh^{2}\theta),\;{\rm for \;} \epsilon=-1,
\end{array}\right.\label{constant}\eeq
where $n$ is the radial quantum number with the following range of definition:
\beq n=\left\{\begin{array}{cc}0,1,\dots
,\infty&{\rm for}\;\epsilon=1\\
0,1,\dots ,n^{max}=[(\delta-j_1-1)/2]&{\rm for}\;\epsilon=-1.
\end{array}\right.
\label{bound}\eeq
The normalization constants are defined
by the expression \beq \frac{ r_0^{2N}
n!\Gamma^2(j_1+1)}{\Gamma(n+j_1+1)} C^2=\left\{
\begin{array}{cc}
(2n+j_1+1+\delta)\Gamma(n+j_1+1+\delta)/\Gamma(n+1+\delta),
&{\rm for}\;\epsilon=1 ,\\
(\delta-2n-j_1-1) \Gamma(\delta-n)/\Gamma (\delta-n-j_1), &{\rm
for}\;\epsilon=-1\;\; .
\end{array}\right.
\label{no}\eeq
 The energy spectrum reads
\beq
\tilde{E}=\left(2n+2j+N+\epsilon\delta\right)^2,
\label{ep}\eeq or, explicitly,
\beq
E_{{n+j},\,s}
=\frac{\epsilon\hbar^2}{2\r}\left[ \left(2n+2j+N+\epsilon\delta\right)^2
-\left(\frac{\omega^2r_0^4}{\hbar^2}+
N^2+{\mu_B}^2\right)
\right].
\label{E0}\eeq
The magnetic flux $\mu_B $ is quantised for  $\epsilon=1 $ and nonquantized
for  $\epsilon=-1$.
 In the flat limit
$\r\rightarrow\infty$ 
 we get the correct formula for the $2N$-dimensional oscillator energy spectrum
\beq E_{\tilde{n},s}=\hbar\sqrt{\omega^2+\frac{B^2_0}{4}}(\tilde{n}+N)+\hbar B_0s,\quad  \tilde{n}=
0,1,2,\ldots,\label{enerF}\eeq
i.e.  ${\tilde n}=2n+2j$ becomes the ``principal'' quantum number.

Thus, we get the following wonderful result: the inclusion of a constant
magnetic field does not change the degeneracy of the  oscillator's spectrum  on
$\DC P^N$ and ${\cal L}_N$.
For $N=1$,
i.e. on the complex projective plane and Lobacewski plane, $s=j$,
hence the spectrum is nondegenerate.
For $N>1$ the spectrum depends
on  $s$ and $\tilde{n}\equiv 2n+2j$, i.e. it is degenerate in
the orbital quantum number $j$.
This degeneracy is due to the existence
of a hidden symmetry.
On the other hand,  for $N=1$
  the complex projective plane/Lobacewski plane coincides
with the sphere/two-sheeted hyperboloid,
while on these spaces there exists an
 oscillator system (Higgs oscillator)
 which possesses a hidden symmetry \cite{higgs}.
However, the inclusion of the constant magnetic field not only breaks
the hidden symmetry (and the degeneracy of the energy spectrum) of that system,
but makes it impossible to get the exact solution of its Schroedinger equation.
So, opposite to the Higgs oscillator case,
the  K\"ahler oscillator  on the  two-dimensional sphere/hyperboloid
 behaves, with respect to the magnetic field, similarly to the planar one.

In free particle limit, i.e. for $\omega=0$, the energy spectrum is described
by the principal quantum number $J$, which
plays the role of the weight of the $SU(N+1)$ group (when
$\epsilon=1$), and the  $SU(N.1)$ group ( when $\epsilon=-1$).
For example, when $\epsilon=1$, the energy spectrum is of the form
\beq
E_{{n+j},\,s}
=\frac{2}{\r}J(J+N)-\frac{\hbar^2}{2\r}\mu_B^2,\qquad
J=n+j+|s+\frac{\mu_B}{2}|,\quad j \geq |s|.
\label{E01}\eeq
As it is seen, the  ground state  becomes degenerate:
the lowest value of $J$ is equal to  $|{\mu_B}|/2$
(see \cite{kn} for details).
Just this  degeneracy plays a key role in the use of quantum mechanics
on $\DC P^N$ in the theory of higher dimensional quantum Hall effect.
For the details we refer to \cite{kn,qheother} and related papers.

\subsubsection*{Conic oscillator}
Our results could be easily extended to the
 family  of $\nu-$ parametric cones (over $\DC P^N$ and ${\cal L}_N$)
 defined by the K\"ahler potential
 \beq
K=\frac{r^2_0}{2\epsilon} \log (1+\epsilon(z\bar z)^\nu),\quad
\nu>0; \quad \epsilon=\pm 1. \label{kahler}\eeq
 The
corresponding metric  and oscillator potential
are  given by the expressions
 \beq
 g_{a\bar b}=
\frac{\nu\r (z\bar z)^{\nu -1}}{2(1+\epsilon (z\bar z)^\nu )}
\left(
 \delta_{a\bar b}-\frac{1-\nu+
\epsilon (z\bar z)^\nu}{z\bar z\;
(1+\epsilon (z\bar z)^\nu)}\bar z^a z^b\right),\qquad
V_{osc}=
\frac{\omega^2 \r}{2}(z\bar{z})^\nu.
 \label{gkf}
\eeq
Introducing
\beq
(z\bar z)^{{\nu}/2}=\frac{\tan[\sqrt{\epsilon}\theta]}{\sqrt{\epsilon}},
\quad
\psi=
\frac{f[\theta]\epsilon^{N/2-1/4}}{(\sin[\sqrt{\epsilon}\theta])^{N-1/2}(\cos[\sqrt{\epsilon}\theta])^{1/2}}
\label{subc}
\eeq
and proceeding in a way completely similar to the previous  case,
 we arrive to the equation (\ref{pt}),
where  $\tilde{E}$
is defined as in (\ref{te}), while the parameters $\delta$ and $j_1$
look as follows:
\beq \delta^2=\frac{\omega^2r_0^4}{\epsilon^2\hbar^2}+
\left(2\frac{s}{\nu}+\frac{B_0\r}{2\epsilon\hbar}\right)^2 \qquad
j^2_1=\frac{(2j+N-1)^2}{\nu}+\frac{\nu-1}{\nu}\left((N-1)^2-
\frac{4s^2}{\nu}\right).
\label{j1} \eeq
Thus, the wavefunctions and energy spectrum
of the conic oscillator are defined, respectively,
 by the expressions (\ref{constant}),(\ref{bound}) and (\ref{ep}).
 The normalization
constants  $C_{cone}$ are
 of the form $C_{cone}=C/\nu^{N/2}$, where $C$ are defined by the
 expressions (\ref{no}).

Explicitly, the energy spectrum of the conic oscillator reads
 \beq
E_{n,\,j,\,s} =\frac{\epsilon\hbar^2}{2\r} \left[(2n+ j_1+1)^2+
\frac{4s^2}{\nu^2}-N^2
\right]+\frac{|\epsilon|\hbar^2}{\r}\delta(2n+ j_1+1)+\hbar
B_0\frac{s}{\nu} \,\label{Eo1}.\eeq

\subsubsection*{KS-transformation}
There is a well-known Kustaanheimo-Stiefel (KS) transformation \cite{ks}
  relating the four-dimensional oscillator with the
three-dimensional Coulomb (and MIC-Kepler \cite{zwanziger}) system.
It allows for a straightforward extension  to the oscillator on
four-dimensional sphere $S^4$ and two-sheet hyperboloids $\DH_{4}$ \cite{np},
 as well as
on the  $\DC P^2$, ${\cal L}_2$ and related cones \cite{bny}.
The KS-transformation of the oscillator on $S^4$, $\DH_4$, $\DC P^2$,
${\cal L}_2$ yields  the MIC-Kepler system  on the three-dimensional
two-sheet hyperboloid. The KS transformation of the oscillator on the
cones over $\DC P^2$ and ${\cal L}_2$ results in the MIC-Kepler system
on the three-dimensional cones over $\DH_3$ equipped with the metric
(see \cite{bny} for  details)
\beq
ds^2_{\nu,  R^2_0}=
\frac{8\nu R^2_0  y^{2\sqrt{\nu}-2}(d{\bf y})^2}{(1- y^{2\sqrt{\nu}})^2},
\qquad R_0=\r.
\label{mred}\eeq
The Hamiltonian of the system is given by the expression
\beq
\widehat{\cal H}_{MIC}=
\frac{1}{\sqrt{g}}\widehat\pi_i{\sqrt{g}}g^{ij}\widehat\pi_j+
 s^2 \hbar^2
\frac{(1- y^{2\sqrt{\nu}} )^2}{8\nu^2 R^2_0 y^{2\sqrt{\nu}}}
-
\frac{\gamma}{2 R_0}\frac{1+ y^{2\sqrt{\nu}}}
{y^{\sqrt{\nu}}},\label{hmic}
\eeq
where
\beq
\widehat{\bp}=-\imath\hbar\frac{\partial}{\partial {\bf y}}-
s{\bf A(y)},\quad [\widehat\pi_i,\widehat\pi_j]=
\hbar s\epsilon_{ijk}\frac{y^k}{y^3}.
\eeq
The coordinates of the initial and final systems are related
as follows:
\beq
{\bf y}=\left({\sqrt{1+\epsilon(z\bar z)^{\nu}}-1}
\right)^{2/\nu}\frac{z\bs\bar z}{(z\bar z)^2}
\eeq
The energy and coupling constant $\gamma$ of this system are defined
by the energy and frequency of the respective four-dimensional oscillator.
The quantum number $s=0,\pm 1/2,1,\ldots$
becomes a fixed parameter (the ``monopole number''), and
 instead of (\ref{jsm}) one  has
\beq
j=|s|,|s|+1,\ldots; \qquad m=-j, -j+1,\ldots , j-1,j.
\eeq
It appears, that applying the KS-transformation to the four-dimensional
oscillator in a constant magnetic field, we have to get the modification
of the MIC-Kepler system on the three-dimensional hyperboloid
(and related cones), which nevertheless
remains superintegrable (exactly solvable).

Surprisingly, repeating the whole procedure, one can find that the inclusion
of the magnetic field in the initial system yields, in the resulting
system, a redefinition of the coupling constant $\gamma$ and the energy
${\cal E}$ only
\beq
\gamma=\frac{E_{\rm osc}}{2}+\frac{\epsilon\hbar^2}{\r}(1-\frac{s^2}{\nu^2})-\frac{B_0}{2\nu}\hbar s ,\quad
 -2{\cal E}=\omega^2+ \frac{\epsilon E_{osc}}{\r} +\frac{\hbar^2}{r^4_0}(1+2\frac{s^2}{\nu^2})+\frac{\epsilon B_0}{\r \nu}\hbar s.
\label{constants}\eeq
Using the expressions (\ref{constant}),
one can   convert the energy spectrum
of the oscillator in the energy spectrum of the
 MIC-Kepler system
\beq
{\cal E}=-
\frac{2\left(\gamma
 -\epsilon\hbar^2(2n+j_1+1)^2/(4R_0)\right)^2}{\hbar^2(2n+j_1+1)^2}-
\frac{\epsilon\gamma}{R_0}
+\frac{\hbar^2}{2 R^2_0},
\eeq
where ${j}_1$ is defined by the expression (\ref{j1}).

\subsubsection*{ Summary and Conclusion}
Let us summarize our results.
We have shown that the inclusion of a constant magnetic field preserves the
hidden symmetries of the oscillator on the complex projective space
$\DC P^N$, in its noncompact version, i. e. the Lobacewski space ${\cal L}_N$.
We constructed  the complete basis of the wavefunction of these systems and
their spectra and found that the inclusion of a constant magnetic field
does not change the qualitative quantum properties of those systems.
Particulary, the inclusion of the magnetic field does not change the
degeneracy of the energy spectra.
These results are  extended to the oscillators on cones related with
$\DC P^N$ and  ${\cal L}_N$.
In some sense, we have  shown that the oscillators on
$\DC P^N$ and  ${\cal L}_N$ (with and without constant magnetic field)
 are more similar to the oscillator
on $\DC^N$ in presence of the constant magnetic field, than to the one in
its absence.
Another  observation concerns the reduction
of the four-dimensional
oscillator and the three-dimensional Coulomb-like system (KS
transformation): we found, to our surprise, that the
oscillators with and without (constant) magnetic fields
result in the equivalent Coulomb-like systems.
Notice, that for the $N=1$ our system remains exactly solvable
in the presence of magnetic field, though it has no hidden symmetries.
In opposite to our model, the well-known Higgs oscillator on
$\DC P^N$ and  ${\cal L}_N$, looses its exact solvability property in the
presence of constant magnetic field, while in it absence it has a hidden
symmetries. So, one can suppose, that considered system would preserve
the exact solvability also in noncommutative case.
Such a modification   seems to be interesting due to the interesting
rotational properties of noncommutative quantum mechanics in the constant
magnetic field, observed at first for the planar case \cite{ncqm}
and later on extended to the two-dimensional sphere
and hyperboloid \cite{ncqms}. While the noncommutative planar oscillator
with constant magnetic field remains superintegrable \cite{npnc},
on the noncommutative spheres and hyperboloids only the particle systems
without potential terms \cite{npnc,npncs}.

\subsubsection*{{ Acknowledgments}}
We are indebted to  Levon Mardoyan and Corneliu Sochichiu for useful
conversations and their interest in this work.
The work of S.B. was supported in part
by the European Community's Human Potential
Programme under
contract HPRN-CT-2000-00131 Quantum Spacetime,
the INTAS-00-00254 grant and the
NATO Collaborative Linkage Grant PST.CLG.979389.
The work of A.N. was supported by grant INTAS 00-00262.

\subsubsection*{Appendix: The choice of the angular coordinates}\label{ap:an}
It is convenient to pass to spherical coordinates
by the use of the so-called ``Smorodinsky's trees method''.
Let us illustrate this method on the  simplest cases of
$N=2$ and $N=3$. Its extension to higher dimensions is straightforward.
 For choosing the appropriate
 angular coordinates we  build Smorodinsky trees (fig.\ref{fig1}).

\begin{figure}[tbph]
\begin{center}
\epsfig{figure=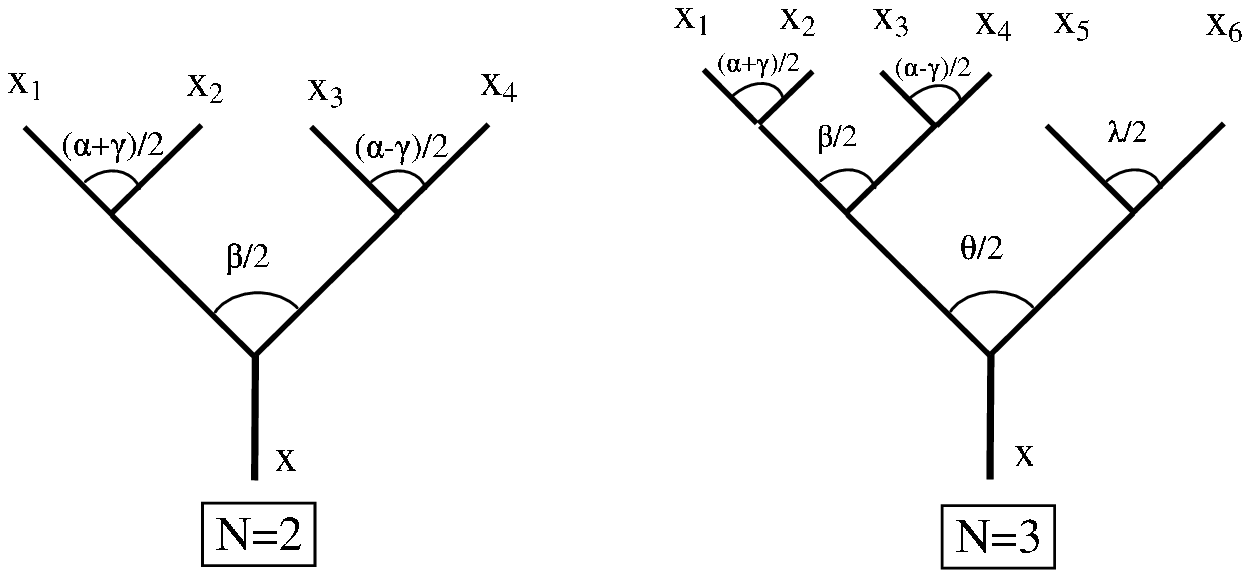}
\end{center}
\caption{Smorodinsky trees for the cases N=2 and N=3.} \label{fig1}
\end{figure}
By means of these trees we express the
Cartesian coordinates $x_1,\ldots, x_{2N}$,
via spherical ones, $x,\phi_i $ by the
following rules.
The  ends of the  top  of the branches mark
 Cartesian coordinates.
 The stock corresponds to the radial coordinate $x$.
Each node marks  some angle $\phi_i$. The angles marked by the top nodes
have the range of definition $[0,2\pi)$,  the remaining angles have the
range of definition $[0,\pi/2)$.
 For the expansion of Cartesian coordinates
in spherical ones   we have to go to each top
 starting from the stock.
When we are passing through a node (denoting the $\phi$ angle) to the left,
we must write $\cos\phi$,
whereas when passing through a node to the right, we  must write  $\sin\phi$.

Explicitly, one has\\
for $N=2$
\begin{eqnarray}
z_1&=&x_1+{i}x_2, \quad z_2=x_3+{i}x_4; \\ \nonumber
x_1&=&x\cos[\beta/2]\cos[(\alpha+\gamma)/2], \quad
x_2=x\cos[\beta/2]\sin[(\alpha+\gamma)/2], \\ \nonumber
x_3&=&x\sin[\beta/2]\cos[(\alpha-\gamma)/2],
 \quad x_4=x\sin[\beta/2]\sin[(\alpha-\gamma)/2]. \\ \nonumber
&&\beta\in[0, \pi),\quad \alpha\in[0, 4\pi),\quad \gamma\in[0, 2\pi)\nonumber
\end{eqnarray}
for $N=3$
\begin{eqnarray}
z_1&=&x_1+{i}x_2, \quad z_2=x_3+{i}x_4, \quad z_3=x_5+{i}x_6; \\ \nonumber
x_1&=&x\cos[\theta/2]\cos[\beta/2]\cos[(\alpha+\gamma)/2],
\quad x_2=x\cos[\theta/2]\cos[\beta/2]\sin[(\alpha+\gamma)/2], \\ \nonumber
x_3&=&x\cos[\theta/2]\sin[\beta/2]\cos[(\alpha-\gamma)/2],
 \quad x_4=x\cos[\theta/2]\sin[\beta/2]\sin[(\alpha-\gamma)/2]. \\ \nonumber
x_5&=&x\sin[\theta/2]\cos[\lambda/2],
\quad x_6=x\sin[\theta/2]\sin[\lambda/2]. \\ \nonumber
&&\beta,\theta\in[0, \pi),\quad \alpha,\lambda\in[0, 4\pi),
\quad \gamma\in[0, 2\pi) \nonumber
\end{eqnarray}
In these coordinates the  operators  $\bf{J}$ an $J_0$
look as follows:\\
for $N=2$
\begin{eqnarray}
\bf{J}^2&=&-\frac{1}{\sin\beta}\frac{\partial}{\partial\beta}\left(\sin\beta\frac{\partial}{\partial\beta}\right)-
 \frac{1}{\sin^2\beta}\left[ \frac{\partial^2}{\partial\alpha^2}+ \frac{\partial^2}{\partial\gamma^2}-
 2 \cos\beta\frac{\partial^2}{\partial\alpha\partial\gamma}\right], \\ \nonumber
J_0&=&-{i}\frac{\partial}{\partial\alpha}, \nonumber
\end{eqnarray}
for $N=3$
\begin{eqnarray}
\bf{J}^2&=&-\frac{1}{\cos^2[\theta/2]}\left[\frac{1}{\sin\beta}\frac{\partial}{\partial\beta}\left(\sin\beta\frac{\partial}{\partial\beta}\right)+
 \frac{1}{\sin^2\beta}\left[ \frac{\partial^2}{\partial\alpha^2}+ \frac{\partial^2}{\partial\gamma^2}-
 2 \cos\beta\frac{\partial^2}{\partial\alpha\partial\gamma}\right]\right]- \\ \nonumber
&-&\frac{1}{\sin^2[\theta/2]}\frac{\partial^2}{\partial\lambda^2}-\frac{1}{\cos^3[\theta/2]\sin[\theta/2]}
\frac{\partial}{\partial\theta}\left(\cos^3[\theta/2]\sin[\theta/2]\frac{\partial}{\partial\theta}\right), \\ \nonumber
J_0&=&-{i}\left[\frac{\partial}{\partial\alpha}+\frac{\partial}{\partial\lambda}\right]. \nonumber
\end{eqnarray}
Without efforts this algorithm could be applied for any finite $N$.


\begin{thebibliography}{99}
\bibitem{higgs}P.~W.~Higgs,
J.\ Phys. {\bf A12} (1979) 309;

  H.~I.~Leemon,
J.\ Phys. {\bf A12} (1979) 489.
\bibitem {pogos}
A.~Barut, A.~Inomata and G.~Junker,~J.~Phys. {\bf A20}~(1987)~6271;
J.~Phys.~{\bf A23}~(1990)~1179;

D.~Bonatos,~C.~Daskaloyanis and K.~Kokkatos,~Phys.~Rev. {\bf A50}~(1994)~3700;

C.~Grosche,~G.~S.~Pogosyan and A.~N.~Sissakian, Fortschritte der Physik,
 {\bf 43}~(6) (1995)~523;

E.~G.~Kalnins, W.~J.~Miller and G.~S.~Pogosyan,
Phys.\ Atom.\ Nucl.\  {\bf 65} (2002) 1086.

\bibitem{cpn}S.~Bellucci and A.~Nersessian,
Phys.\ Rev. {\bf D67} (2003) 065013 .
\bibitem{ny}
A.~Nersessian and A.~Yeranyan,
J.\ Phys.\  {\bf A37} (2004) 2791.
\bibitem{bny}S.~Bellucci,A.~Nersessian and A.~Yeranyan,
``Quantum Mechanics Model on Kaehler conifold''
arXiv: hep-th/0312323 , to appear in Phys.\ Rev. {\bf D70} (15 July, 2004) issue 2.
\bibitem{sqs}
S.~Bellucci and A.~Nersessian,
``Supersymmetric Kaehler oscillator in a constant magnetic field,''
arXiv:hep-th/0401232.
\bibitem{zhang} S.~C.~Zhang and J.~P.~Hu,
Science {\bf 294} (2001) 823.
\bibitem{kn}D.~Karabali and  V.~P.~Nair,
 Nucl.\ Phys. {\bf B641} (2002) 533.

\bibitem{qheother}B.~A.~Bernevig, J.~P.~Hu, N.~Toumbas and S.~C.~Zhang,
Phys.\ Rev.\ Lett.\  {\bf 91} (2003) 236803;

M.~Fabinger,
JHEP {\bf 0205} (2002) 037;

S.~Bellucci, P.~Y.~Casteill and A.~Nersessian,
Phys.\ Lett.\  {\bf B574} (2003) 121;

D.~Karabali and V.~P.~Nair,
Nucl.\ Phys.\  {\bf B679} (2004) 427;

K.~Hasebe and Y.~Kimura,
``Dimensional hierarchy in quantum Hall effects on fuzzy spheres'',
arXiv:hep-th/0310274;

D.~Karabali and V.~P.~Nair,
``Edge states for quantum Hall droplets in higher dimensions and a generalized
arXiv:hep-th/0403111.

\bibitem{ks}P.~Kustaanheimo and~E.~Stiefel,~J.~Reine Angew Math.
 {\bf 218} (1965) 204;

see, also the review
V.~Ter-Antonyan, {\sl Dyon-Oscillator Duality}, arXiv:quant-ph 0003106.

\bibitem{zwanziger}D.~Zwanziger, Phys.\ Rev. {\bf 176} (1968) 1480;

 H.~V.~McIntosh and A.~Cisneros, J.\ Math.\ Phys. {\bf 11} (1970) 896.

\bibitem{np}A.~Nersessian and G.~Pogosyan,
Phys.\ Rev. {\bf A63} (2001) 020103(R).
\bibitem{ncqm}
S.~Bellucci, A.~Nersessian and C.~Sochichiu,
Phys.\ Lett.\  {\bf B522} (2001) 345.
\bibitem{ncqms}
S.~Bellucci and A.~Nersessian,
Phys.\ Lett.\  {\bf B542} (2002) 295.
\bibitem{npnc}
V.~P.~Nair and A.~P.~Polychronakos,
Phys.\ Lett.\  {\bf B505} (2001) 267.

\bibitem{npncs}
R.~Iengo and R.~Ramachandran,
JHEP {\bf 0202} (2002) 017;

D.~Karabali, V.~P.~Nair and A.~P.~Polychronakos,
Nucl.\ Phys.  {\bf B627} (2002) 565.
\end{thebibliography}
\end{document}